%% file: cpu-eval-date2027-camera-ready.tex
\documentclass[conference]{IEEEtran}

\usepackage{cite}
\usepackage{amsmath}
\usepackage{booktabs}
\usepackage{array}
\usepackage{multirow}
\usepackage{graphicx}
\usepackage{xurl}
\usepackage{placeins}
\usepackage{dblfloatfix}
\usepackage[a4paper,total={184mm,239mm}]{geometry}

\input{problem-grid.tex}
\newcommand{\FiveOllamaCVA}{32.04 & 45.16 & 29.78 & 45.16 & 25.59 & 43.01 & 34.73 & 56.99 & 36.13 & 55.91}
\newcommand{\GPTFiveSixSolCVA}{56.34 & 69.89}
\newcommand{\ClaudeOpusCVA}{73.87 & 90.32}
\newcommand{\SevenModelMeanCVA}{41.21}
\newcommand{\SevenModelMeanPassTenCVA}{58.06}
\newcommand{\FiveOllamaIbex}{24.50 & 30.00 & 22.50 & 30.00 & 16.50 & 30.00 & 26.00 & 40.00 & 25.00 & 35.00}
\newcommand{\GPTFiveSixSolIbex}{47.50 & 60.00}
\newcommand{\ClaudeOpusIbex}{68.00 & 90.00}
\newcommand{\SevenModelMeanIbex}{32.86}
\newcommand{\SevenModelMeanPassTenIbex}{45.00}
\newcommand{\FiveOllamaSweRVELTwo}{37.32 & 57.75 & 38.87 & 57.75 & 35.77 & 50.70 & 37.89 & 63.38 & 39.72 & 60.56}
\newcommand{\GPTFiveSixSolSweRVELTwo}{62.39 & 78.87}
\newcommand{\ClaudeOpusSweRVELTwo}{69.72 & 85.92}
\newcommand{\SevenModelMeanSweRVELTwo}{45.96}
\newcommand{\SevenModelMeanPassTenSweRVELTwo}{64.99}
\newcommand{\FiveOllamaRiftTwoCore}{9.58 & 16.67 & 8.33 & 27.08 & 10.83 & 18.75 & 18.33 & 37.50 & 18.96 & 39.58}
\newcommand{\GPTFiveSixSolRiftTwoCore}{52.08 & 70.83}
\newcommand{\ClaudeOpusRiftTwoCore}{56.04 & 79.17}
\newcommand{\SevenModelMeanRiftTwoCore}{24.88}
\newcommand{\SevenModelMeanPassTenRiftTwoCore}{41.37}
\newcommand{\FiveOllamaVexRiscv}{0.00 & 0.00 & 1.25 & 6.25 & 0.00 & 0.00 & 2.50 & 12.50 & 5.00 & 6.25}
\newcommand{\GPTFiveSixSolVexRiscv}{10.63 & 18.75}
\newcommand{\ClaudeOpusVexRiscv}{14.37 & 56.25}
\newcommand{\SevenModelMeanVexRiscv}{4.82}
\newcommand{\SevenModelMeanPassTenVexRiscv}{14.29}
\newcommand{\FiveOllamaBOOM}{2.75 & 6.59 & 5.93 & 16.48 & 6.26 & 14.29 & 10.66 & 20.88 & 9.67 & 21.98}
\newcommand{\GPTFiveSixSolBOOM}{30.77 & 49.45}
\newcommand{\ClaudeOpusBOOM}{57.25 & 78.02}
\newcommand{\SevenModelMeanBOOM}{17.61}
\newcommand{\SevenModelMeanPassTenBOOM}{29.67}
\newcommand{\FiveOllamaAllCPUEval}{20.15 & 30.38 & 20.47 & 34.81 & 18.70 & 30.68 & 24.57 & 42.77 & 25.22 & 41.89}
\newcommand{\GPTFiveSixSolAllCPUEval}{47.46 & 63.42}
\newcommand{\ClaudeOpusAllCPUEval}{62.86 & 82.89}
\newcommand{\SevenModelMeanAllCPUEval}{31.35}
\newcommand{\SevenModelMeanPassTenAllCPUEval}{46.69}
\newcommand{\FiveOllamaVerilogEvalVTwo}{69.23 & 76.92 & 69.55 & 83.97 & 69.17 & 82.05 & 74.74 & 85.26 & 69.81 & 84.62}
\newcommand{\GPTFiveSixSolVerilogEvalVTwo}{90.19 & 94.23}
\newcommand{\ClaudeOpusVerilogEvalVTwo}{92.69 & 95.51}
\newcommand{\SevenModelMeanVerilogEvalVTwo}{76.48}
\newcommand{\SevenModelMeanPassTenVerilogEvalVTwo}{86.08}
\newcommand{\SevenModelDifferenceRow}{\textminus49.08 & \textminus46.54 & \textminus49.08 & \textminus49.17 & \textminus50.46 & \textminus51.37 & \textminus50.17 & \textminus42.48 & \textminus44.59 & \textminus42.73 & \textminus42.73 & \textminus30.81 & \textminus29.83 & \textminus12.62 & \textbf{\textminus45.14} & \textminus39.39}
\newcommand{\OpusCPUOverallPassOne}{62.86}
\newcommand{\OpusVEOverallPassOne}{92.69}
\newcommand{\SevenModelMeanPassOneGap}{45.14}
\newcommand{\SourceSummaryBOOM}{91 & 82 & 171.0 & 15,557}
\newcommand{\SourceSummaryCVA}{93 & 152 & 368.0 & 34,224}
\newcommand{\SourceSummaryRiftTwoCore}{48 & 38.5 & 75.2 & 3,610}
\newcommand{\SourceSummarySweRVELTwo}{71 & 122 & 328.2 & 23,299}
\newcommand{\SourceSummaryVexRiscv}{16 & 170 & 275.0 & 4,400}
\newcommand{\SourceSummaryIbex}{20 & 679.5 & 816.8 & 16,336}
\newcommand{\SourceSummaryOverall}{339 & 106 & 287.4 & 97,426}
\renewcommand{\InstructionsColumnWidth}{6.3mm}
\begin{document}

\title{CPU-Eval: Evaluating Large Language Models\\for CPU Hardware Code Generation}
\author{\IEEEauthorblockN{Yuisei Iwata\IEEEauthorrefmark{1}, Yutaka Masuda\IEEEauthorrefmark{2}, and Tohru Ishihara\IEEEauthorrefmark{2}}
\IEEEauthorblockA{\IEEEauthorrefmark{1}School of Informatics, Nagoya University, Nagoya, Japan\\
\IEEEauthorrefmark{2}Graduate School of Informatics, Nagoya University, Nagoya, Japan\\
Email: yiwata@nagoya-u.jp, \{masuda, ishihara\}@ertl.jp}}
\maketitle

\begin{abstract}
Benchmarks such as VerilogEval have supported rapid advances in hardware description language (HDL) generation by large language models (LLMs).
However, recent HDL-generation systems achieve pass rates above 90\% on VerilogEval, leaving limited headroom for revealing further advances.
To provide a challenging target for further progress on a practically important hardware-development task, we introduce CPU-Eval, a benchmark of 339 HDL-generation problems constructed from six CPUs.
Drawn from practical CPU designs, including families fabricated in silicon, the problems span CPUs from small embedded cores to Linux-capable out-of-order cores and cover diverse CPU functions written in SystemVerilog, Chisel, and SpinalHDL.
In each problem, an LLM performs the real CPU-development task of generating HDL.
Across seven state-of-the-art LLMs, mean pass@1 is \SevenModelMeanPassOneGap{} percentage points lower on CPU-Eval than on VerilogEval v2 Spec-to-RTL, exposing capability differences obscured by VerilogEval's high pass rates.
CPU-Eval reveals that pass@1 is lower for submodules with more source-code lines, which motivates fine-grained modular design for LLM-aided hardware generation.
We also show that pass@1 is lower for CPUs written in less widely used HDLs, motivating designers to verify that an LLM is sufficiently proficient at generating code in the HDL they use before deployment.
\end{abstract}

\section{Introduction}
\begin{figure*}[!b]
  \centering
  \includegraphics[width=\textwidth]{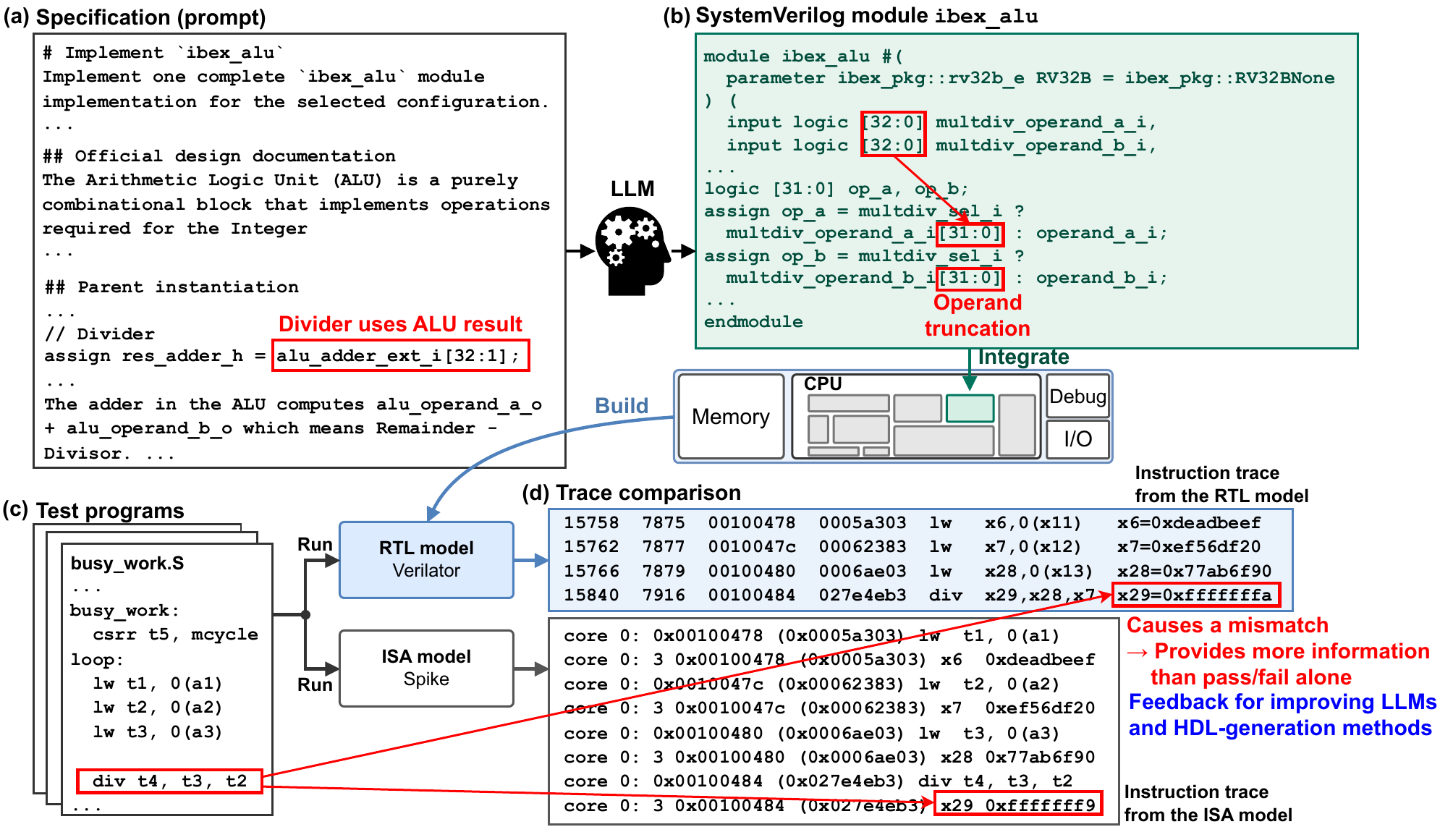}
  \caption{Generation and evaluation process for one of the 339 CPU-Eval problems: (a) CPU-Eval provides the prompt for the arithmetic logic unit (ALU) of the Ibex CPU to an LLM, (b) the LLM generates \texttt{ibex\_alu} and CPU-Eval integrates it into Ibex, (c) CPU-Eval executes a test program, and (d) trace comparison locates a functional error.}
  \label{fig:example}
\end{figure*}
Benchmarks such as VerilogEval~\cite{verilogeval,verilogeval2} have supported rapid advances in hardware description language (HDL) generation by large language models (LLMs).
Benchmarks allow researchers to evaluate whether new LLMs and generation methods produce more functionally correct HDL, while compilation and functional-test failures provide concrete feedback for improving models and developing generation techniques.
At training time, VeriReason extended the RTLCoder dataset with reasoning steps and testbenches, used testbench results as rewards for reinforcement learning, and used VerilogEval to show that the resulting models generated more functionally correct RTL~\cite{rtlcoder,verireason}.
At inference time, RTLFixer used compiler errors from VerilogEval problems to iteratively repair syntax errors~\cite{rtlfixer}, while VerilogCoder used VerilogEval testbench failures and waveform tracing to locate and repair functional errors~\cite{verilogcoder}.
Benchmarks therefore provide a common basis for analyzing failures and developing models and generation methods, beyond simply comparing them.

For this research cycle to continue, a benchmark must reveal measurable differences among HDL-generation approaches and provide challenging cases for failure analysis and repair.
However, recent systems already achieve pass rates above 90\% on several VerilogEval problem sets~\cite{mage,hdlforge,verirefine}, leaving fewer unresolved cases for further investigation.
Some benchmarks address this issue with challenging problems drawn from open-source hardware designs~\cite{realbench,chipbench}.
RealBench, for example, asks LLMs to generate modules from four open-source hardware IPs and reports that o1-preview achieves a pass@1 of 13.3\%, compared with its 70.5\% on VerilogEval v2 Spec-to-RTL (S2R)~\cite{realbench}.
Still, existing benchmarks do not adequately measure how well LLMs can implement submodules within complete CPUs because they cover only a limited range of CPU designs and do not evaluate generated submodules as part of those CPUs.

To address this gap, we introduce CPU-Eval, a 339-problem benchmark for generating submodules within six complete, practical CPUs.
Thanks to its broad CPU coverage, CPU-Eval yields practical insights for LLM-aided processor design, as detailed in Section~\ref{sec:experiments}.
CPU-Eval's task reflects how designers commonly build on an existing CPU and design only the additional submodules needed for a particular purpose rather than designing an entire CPU from scratch~\cite{atasu2003automatic,yu2004scalable}.

We make the following contributions:
\begin{enumerate}
  \item We construct 339 HDL-generation problems from six practical CPU implementations, including CPU families fabricated in silicon.
  The problems range from small embedded cores to Linux-capable out-of-order cores and cover diverse CPU functions in datapaths and complex control logic.
  CPU-Eval\footnote{CPU-Eval is available at \url{https://github.com/next-edge-ai-ishihara/cpu-eval}.} covers three HDLs: SystemVerilog at RTL and the higher-level hardware construction languages Chisel and SpinalHDL.
  \item We define a benchmark task for LLMs that combines submodule-level HDL generation with CPU-level functional evaluation without relying on commercial EDA tools.
  CPU-Eval integrates each generated submodule into its CPU, executes test programs on that CPU, and compares the CPU's execution trace with a reference ISA trace to detect deviations from expected CPU behavior.
  \item We reveal differences in model capability hidden by VerilogEval's high pass rates by evaluating seven state-of-the-art LLMs on CPU-Eval, which yields a mean pass@1 \SevenModelMeanPassOneGap{} percentage points lower than that on VerilogEval v2 S2R.
  CPU-Eval shows that pass@1 decreases for submodules with more source-code lines and is lower for CPUs written in less widely used HDLs.
\end{enumerate}

Our findings have two implications for LLM-aided hardware design.
First, the source-code-length trend motivates fine-grained modular design with smaller submodules for LLMs.
Second, the differences among HDLs motivate designers to use practical benchmarks such as CPU-Eval to verify that an LLM is sufficiently proficient at generating code in the HDL they use before deployment.

\section{CPU-Eval Benchmark Design}

\subsection{An Illustrative Example of HDL Generation and Evaluation}

Each CPU-Eval problem pairs a prompt specifying one CPU submodule with a functional test of the generated implementation.
CPU-Eval measures whether HDL generated by LLMs functions correctly within a complete CPU and, when it does not, records where program execution diverges from a reference ISA model.

Fig.~\ref{fig:example} illustrates how an LLM solves and CPU-Eval evaluates one of the 339 problems.
The prompt in Fig.~\ref{fig:example}(a) specifies the arithmetic logic unit (ALU) of the Ibex CPU, and the LLM Gemma~4~31B generates the \texttt{ibex\_alu} submodule shown in Fig.~\ref{fig:example}(b).
The generated HDL incorrectly discards the most significant bit of each divider operand before computing the result used by the divider.
At each division step, Ibex uses this result to decide whether to subtract the divisor and set the corresponding quotient bit, so truncating its operands can change the quotient.

CPU-Eval integrates this HDL into Ibex and builds an RTL model.
It then runs the test program \texttt{busy\_work.S}, shown in Fig.~\ref{fig:example}(c), on both the RTL and Spike ISA models of the CPU.
Fig.~\ref{fig:example}(d) compares the two execution traces.
Immediately before the \texttt{div} instruction, both traces show that \texttt{x28} contains \texttt{0x77ab6f90} (2,007,723,920) and \texttt{x7} contains \texttt{0xef56df20} ($-279{,}519{,}456$), so the correct signed quotient is \texttt{0xfffffff9} ($-7$).
Spike then writes this value to \texttt{x29}, whereas operand truncation makes the RTL model write \texttt{0xfffffffa} ($-6$).
The traces thus identify the instruction and architectural-state update that expose the functional error, rather than reporting only that the generated HDL fails.
Researchers can analyze such concrete failures to improve LLMs and HDL-generation methods.

\subsection{Scale and Diversity}

\begin{table*}[!t]
  \centering
  \caption{Six open-source RISC-V CPUs in CPU-Eval and their characteristics.}
  \label{tab:cpus}
  \setlength{\tabcolsep}{0pt}
  \renewcommand{\arraystretch}{1.0}
  \footnotesize
  \begin{tabular*}{\textwidth}{@{}l@{\hspace{6pt}}l@{\extracolsep{\fill}}l@{\extracolsep{0pt}}r@{\hspace{7pt}}r@{\hspace{7pt}}r@{\hspace{7pt}}r@{}}
    \toprule
    CPU & Description & HDL & \multirow{2}{*}{\shortstack{No. of \\ submodules}} & \multicolumn{3}{c}{Source-code lines} \\
    \cmidrule(l){5-7}
    & & & & Median & Mean & Total \\
    \midrule
    BOOM & 64-bit out-of-order core with 3-wide decode, a 96-entry reorder buffer, and Linux support & Chisel & \SourceSummaryBOOM \\
    CVA6 & 32/64-bit in-order core with a 6-stage pipeline and Linux-capable configurations & SystemVerilog & \SourceSummaryCVA \\
    Rift2Core & 64-bit out-of-order core with a 9-stage pipeline and machine, supervisor, and user modes & Chisel & \SourceSummaryRiftTwoCore \\
    VeeR EL2 & 32-bit in-order core with a 4-stage pipeline, branch prediction, and machine/user modes & SystemVerilog & \SourceSummarySweRVELTwo \\
    VexRiscv & 32-bit in-order core optimized for FPGAs, with optional caches, an MMU, and Linux support & SpinalHDL & \SourceSummaryVexRiscv \\
    Ibex & 32-bit in-order core with a 2-stage pipeline, suited for embedded control applications & SystemVerilog & \SourceSummaryIbex \\
    \midrule
    Overall & & & \SourceSummaryOverall \\
    \bottomrule
  \end{tabular*}
\end{table*}

\begin{table*}[!t]
  \setlength{\belowcaptionskip}{2pt}
  \caption{%
    The 91 HDL-generation problems for BOOM submodules and their source-code lines.
    Insn. is the number of instructions compared between execution traces.
    $\dagger$ indicates an additional check not recorded in those traces.}
  \label{tab:boom-problems}
  \vspace{-2.5ex}
  \centering
  \setlength{\tabcolsep}{0pt}
  \renewcommand{\arraystretch}{0.45}
  \renewcommand{\InstructionsColumnWidth}{7.8mm}
  \renewcommand{\ProblemNameLeading}{6.6}
  \renewcommand{\ProblemRowGapExtra}{0.5pt}
  \renewcommand{\ProblemNameVAlign}{b}
  \renewcommand{\ProblemValueVAlign}{b}
  \renewcommand{\ProblemNameEndStrut}{\strut}
  \renewcommand{\ProblemLineSkipLimit}{-1000pt}
  \renewcommand{\SetProblemValueShift}{%
    \setlength{\ProblemValueShift}{\ht\ProblemNameBox}%
    \addtolength{\ProblemValueShift}{\dp\ProblemNameBox}%
    \addtolength{\ProblemValueShift}{-\ht\ProblemSingleLineBox}%
    \addtolength{\ProblemValueShift}{-\dp\ProblemSingleLineBox}%
    \setlength{\ProblemValueShift}{0.5\ProblemValueShift}}
  \ProblemGrid{\BOOMRows}
\end{table*}

CPU-Eval contains 339 HDL-generation problems from six RISC-V CPUs spanning a broad range of CPU designs.
From the Core(s) category of RISC-V Exchange, maintained by RISC-V International~\cite{riscv-exchange-core,riscv-exchange-cores}, we selected CPUs with redistributable HDL and documented verification procedures that use no commercial EDA tools.
As summarized in Table~\ref{tab:cpus}, the CPUs range from small embedded cores to Linux-capable out-of-order cores, and their hardware source is written in SystemVerilog, Chisel, and SpinalHDL.
CPU-Eval targets practical CPU families used to build physical processors, rather than designs created only as benchmark exercises.
For example, Ibex has undergone multiple tape-outs~\cite{ibexrepo}, CVA6's predecessor Ariane was fabricated in 22-nm FDSOI~\cite{zaruba2019cva6}, a BOOM-based core was fabricated in 28-nm CMOS~\cite{celio2019broom}, and VexRiscv was used in the taped-out Baochip 1x~\cite{baochip1x}.
CPU-Eval is publicly available, allowing researchers to evaluate and develop HDL-generation methods without access to commercial EDA tools.

For each CPU, we enumerated every submodule that forms part of the CPU in the selected configuration.
Table~\ref{tab:boom-problems} lists all 91 BOOM problems as one example of this enumeration.
This example illustrates the broad range of CPU functions covered by these submodules, including arithmetic units, instruction decoding, pipeline control, forwarding, branch prediction, caching, and memory management.

Table~\ref{tab:cpus} reports, for each CPU, the median and mean number of source-code lines per submodule and the total across all submodules.
In Table~\ref{tab:boom-problems}, the Lines column reports the source-code length for each BOOM submodule.
The Insn. column reports the number of retired instructions compared between the RTL and ISA models across the test programs, with ``k'' denoting 1,000.

A dagger marks problems for which CPU-Eval checks CPU behavior that is not recorded in the execution traces.
For example, the test program for \texttt{LoopBranchPredictorBank} writes its measured cycle count to \texttt{x20}, which CPU-Eval compares with the expected value.
The page-table walker (PTW), which traverses page tables after a TLB miss, is evaluated by supervisor-level programs that check dirty-bit handling and instruction fetches after page remapping.
The directed test for \texttt{BoomRAS} writes return addresses into the stack, reads them by index, and compares each \texttt{read\_addr} output with the expected address.
These additional checks allow CPU-Eval to include all 91 BOOM submodules, including important functions such as branch prediction and virtual-memory translation that cannot be evaluated from retired-instruction traces alone.

\begin{table*}[!t]
  \centering
  \caption{Problem counts, included CPU designs, tasks, and evaluation procedures in CPU-Eval and related hardware-code benchmarks.}
  \label{tab:benchmark-comparison}
  \setlength{\tabcolsep}{3pt}
  \renewcommand{\arraystretch}{1.08}
  \footnotesize
  \begin{tabular*}{\textwidth}{@{}l@{\extracolsep{\fill}}r >{\raggedright\arraybackslash}p{0.22\textwidth}>{\raggedright\arraybackslash}p{0.22\textwidth}>{\raggedright\arraybackslash}p{0.29\textwidth}@{}}
    \toprule
    Benchmark & No. of problems & CPU designs included & What the LLM does & How the benchmark evaluates the output \\
    \midrule
    VerilogEval v2~\cite{verilogeval2}
      & 156
      & None. Small standalone modules from educational exercises
      & Generate one RTL module from a specification
      & Simulate using the testbench provided for each problem \\ \addlinespace[2pt]
    RTLLM~\cite{rtllm}
      & 30
      & One 518-line simplified RISC CPU
      & Generate all RTL for one design from a specification
      & Simulate using the testbench provided for each problem \\ \addlinespace[2pt]
    RTLLM 2.0~\cite{openllmrtl}
      & 50
      & Small standalone modules such as a 32-bit MIPS ALU
      & Generate one RTL module from a specification
      & Simulate using the testbench provided for each problem \\ \addlinespace[2pt]
    RealBench~\cite{realbench}
      & 64
      & One 32-bit embedded RISC-V CPU
      & Generate one module or an entire design from a specification
      & Use testbench simulation and formal equivalence checking \\ \addlinespace[2pt]
    ChipBench~\cite{chipbench}
      & 265
      & Nine modules from open-source CPUs
      & Generate or debug Verilog, Python, SystemC, or CXXRTL
      & Compare the generated output behavior with the golden Verilog implementation \\ \addlinespace[2pt]
    ChipVerilog~\cite{chipverilog}
      & 64
      & Two small RISC cores: OR1200 and a five-stage MIPS-16 core
      & Generate one module or an entire IP/core from a specification
      & Use testbench simulation or formal equivalence checking \\ \addlinespace[2pt]
    HWE-Bench~\cite{hwebench}
      & 417
      & Five CPUs in tasks: Ibex, CVA6, Rocket, VeeR EL2, and XiangShan
      & Repair a reported bug in a hardware repository
      & Apply the generated patch and run a task-specific test \\ \addlinespace[2pt]
    \textbf{CPU-Eval}
      & \textbf{339}
      & \textbf{Six CPUs, from embedded to Linux-capable out-of-order cores}
      & \textbf{Generate one CPU submodule from a specification}
      & \textbf{Integrate the generated submodule into the CPU and execute programs on the CPU} \\
    \bottomrule
  \end{tabular*}
\end{table*}

\subsection{Comparison with Existing Benchmarks}

Table~\ref{tab:benchmark-comparison} compares CPU-Eval with existing benchmarks in terms of the number of problems, included CPU designs, generation tasks, and evaluation methods.
Unlike the standalone modules drawn from educational exercises in VerilogEval v2~\cite{verilogeval2} and the hand-crafted standalone designs in RTLLM 2.0~\cite{openllmrtl}, CPU-Eval constructs problems from submodules of complete, practical CPU implementations, including CPU families fabricated in silicon.
CPU-Eval's problem set is more than twice the size of VerilogEval v2 S2R, and the submodules in its problems have substantially more source-code lines: a median of 106 and a mean of 287, versus 16 and 19, respectively.

RTLLM, RealBench, ChipBench, and ChipVerilog do include CPU designs, but their tasks ask an LLM to generate either an entire small CPU or individual modules, and their evaluations do not integrate a generated submodule into a complete CPU and execute programs on it~\cite{rtllm,realbench,chipbench,chipverilog}.
CPU-Eval instead asks an LLM to generate a submodule of a practical CPU and covers six CPUs ranging from small embedded cores to Linux-capable out-of-order cores.

CPU-Eval and HWE-Bench both cover practical CPU designs, but their tasks differ: HWE-Bench asks an LLM agent to repair a reported bug in GitHub repositories~\cite{hwebench}, whereas CPU-Eval asks an LLM to generate HDL for a submodule from its specification.

\begin{table*}[!t]
  \caption{%
    Pass@$k$ (\%) for $k\in\{1,10\}$ on CPU-Eval overall, each CPU, and VerilogEval v2 S2R.
    The signed differences in the final row are CPU-Eval minus VerilogEval v2 in percentage points.}
  \label{tab:results}
  \centering
  \setlength{\tabcolsep}{1.5pt}
  \begin{tabular}{@{}l*{16}{>{\raggedleft\arraybackslash\rightskip=4pt}p{26.6pt}}@{}}
    \toprule
    & \multicolumn{2}{c}{\makebox[0pt][c]{\footnotesize Gemma 4 31B}}
    & \multicolumn{2}{c}{\makebox[0pt][c]{\footnotesize gpt-oss-120B}}
    & \multicolumn{2}{c}{\makebox[0pt][c]{\footnotesize Qwen3.5-397B}}
    & \multicolumn{2}{c}{\makebox[0pt][c]{\scalebox{0.84}[1]{\footnotesize DeepSeek-V4.1-Flash}}}
    & \multicolumn{2}{c}{\makebox[0pt][c]{\footnotesize Kimi K2.7 Code}}
    & \multicolumn{2}{c}{\makebox[0pt][c]{\footnotesize GPT-5.6 Sol}}
    & \multicolumn{2}{c}{\makebox[0pt][c]{\footnotesize Claude Opus 5}}
    & \multicolumn{2}{c}{\makebox[0pt][c]{\footnotesize Mean (7 LLMs)}} \\
    \cmidrule(lr){2-3}\cmidrule(lr){4-5}\cmidrule(lr){6-7}\cmidrule(lr){8-9}
    \cmidrule(lr){10-11}\cmidrule(lr){12-13}\cmidrule(lr){14-15}\cmidrule(l){16-17}
    Problem set
    & $k\!=\!1$ & $k\!=\!10$
    & $k\!=\!1$ & $k\!=\!10$
    & $k\!=\!1$ & $k\!=\!10$
    & $k\!=\!1$ & $k\!=\!10$
    & $k\!=\!1$ & $k\!=\!10$
    & $k\!=\!1$ & $k\!=\!10$
    & $k\!=\!1$ & $k\!=\!10$
    & $k\!=\!1$ & $k\!=\!10$ \\
    \midrule
    VerilogEval v2 & \FiveOllamaVerilogEvalVTwo & \GPTFiveSixSolVerilogEvalVTwo & \ClaudeOpusVerilogEvalVTwo & \SevenModelMeanVerilogEvalVTwo & \SevenModelMeanPassTenVerilogEvalVTwo \\
    \addlinespace
    CVA6 & \FiveOllamaCVA & \GPTFiveSixSolCVA & \ClaudeOpusCVA & \SevenModelMeanCVA & \SevenModelMeanPassTenCVA \\
    Ibex & \FiveOllamaIbex & \GPTFiveSixSolIbex & \ClaudeOpusIbex & \SevenModelMeanIbex & \SevenModelMeanPassTenIbex \\
    VeeR EL2 & \FiveOllamaSweRVELTwo & \GPTFiveSixSolSweRVELTwo & \ClaudeOpusSweRVELTwo & \SevenModelMeanSweRVELTwo & \SevenModelMeanPassTenSweRVELTwo \\
    Rift2Core & \FiveOllamaRiftTwoCore & \GPTFiveSixSolRiftTwoCore & \ClaudeOpusRiftTwoCore & \SevenModelMeanRiftTwoCore & \SevenModelMeanPassTenRiftTwoCore \\
    VexRiscv & \FiveOllamaVexRiscv & \GPTFiveSixSolVexRiscv & \ClaudeOpusVexRiscv & \SevenModelMeanVexRiscv & \SevenModelMeanPassTenVexRiscv \\
    BOOM & \FiveOllamaBOOM & \GPTFiveSixSolBOOM & \ClaudeOpusBOOM & \SevenModelMeanBOOM & \SevenModelMeanPassTenBOOM \\
    \cmidrule(lr){1-17}
    All CPU-Eval & \FiveOllamaAllCPUEval & \GPTFiveSixSolAllCPUEval & \ClaudeOpusAllCPUEval & \SevenModelMeanAllCPUEval & \SevenModelMeanPassTenAllCPUEval \\
    Difference (pp) & \SevenModelDifferenceRow \\
    \bottomrule
  \end{tabular}
\end{table*}

\subsection{Problem Quality Checks}
\label{sec:benchmark-quality}

A low pass rate alone does not establish that an HDL-generation problem is difficult because incomplete prompts and defective tests can also cause generated HDL to fail.
For example, RTL-BenchMT found VerilogEval prompts that omit register initialization required by the test~\cite{rtlbenchmt}.
STG found that a flawed VerilogEval testbench can reject correct HDL~\cite{stg}.
We therefore check every CPU-Eval prompt for missing requirements and every test for incorrect verdicts before interpreting an LLM failure as evidence of problem difficulty.

\subsubsection{Checking Prompts for Missing Specifications}

To prevent incomplete specifications, we enumerate the interface and externally required behavior of the submodule in each problem by inspecting the submodule's source code and the places where the CPU connects to it.
We then map every item to verbatim excerpts from design documents, instruction-set and interface specifications, or surrounding HDL.
The resulting prompt provides the design information that a human developer would use to implement the submodule.

\subsubsection{Checking Test Verdicts}

As a reliable starting point for functional evaluation, CPU-Eval uses test programs that CPU developers run to verify their complete CPUs.
For CVA6, CPU-Eval uses the RISC-V Architectural Tests, which check implementations of the RISC-V specification and which the CVA6 developers include in their regression tests~\cite{riscvarchtest,cva6repo}.
The VeeR EL2 evaluation uses the \texttt{riscv-dv} regression tests maintained by the Linux Foundation-hosted CHIPS Alliance and run by the CPU developers in continuous integration~\cite{chipsalliance,riscvdv,swervrepo}.
Rift2Core and BOOM are tested with the \texttt{riscv-tests} processor test suite used by both projects~\cite{riscvtests,rift2corerepo,boomrepo}.
The VexRiscv test programs come from the project's \texttt{TestIndividualFeatures} regression suite~\cite{vexriscvrepo}.
For Ibex, CPU-Eval uses \texttt{simple\_system} programs that the Ibex project runs in continuous integration~\cite{ibexrepo}.

We do not assume that reusing these test programs guarantees a correct verdict for every CPU-Eval problem.
Before including a problem, we first confirm that its tests accept the source code for the submodule in the CPU repository.
We then prepare several HDL implementations, each containing a different functional error while still building and simulating successfully.
The tests must reject each implementation because its execution differs from the expected CPU behavior.
If an erroneous implementation passes, we determine whether the programs failed to exercise the error or the comparison omitted its effect, and strengthen the programs or comparison accordingly.
We repeat these checks after every change.

For example, in the CVA6 \texttt{store\_unit} problem, an implementation that wrote zero for every store initially passed because the comparison recorded instructions and register updates but omitted memory updates.
We therefore extended the comparison to include memory-write information, such as addresses and data values.
The extended comparison rejects both the implementation that writes zero for every store and another that suppresses every memory write, while continuing to accept the \texttt{store\_unit} source code in the CVA6 repository.
In this way, we improve the reliability of CPU-Eval's tests.

\section{Experiments}
\label{sec:experiments}

\subsection{Experimental Setup}

To uncover practical insights for LLM-aided processor design, we evaluate seven state-of-the-art LLMs on CPU-Eval's 339 problems and VerilogEval v2 S2R.
We evaluate five open-weight LLMs served by Ollama Cloud: Gemma 4 31B, gpt-oss-120B, Qwen3.5-397B, DeepSeek-V4.1-Flash, and Kimi K2.7 Code.
We also evaluate two flagship commercial LLMs: GPT-5.6 Sol and Claude Opus 5.
For each problem, we give each LLM the prompt and independently sample 10 HDL implementations.
For the five Ollama Cloud models, we set temperature to 0.8 and top-$p$ to 0.95, matching VerilogEval's sampling settings~\cite{verilogeval}.
For these five models, we set a 65,536-token limit that remains sufficient to generate the HDL required by every CPU-Eval problem while stopping buggy generations that repeat output indefinitely.
We set \texttt{think} to \texttt{low} for gpt-oss-120B and to \texttt{false} for the other four open-weight models, and set the reasoning effort to low for GPT-5.6 Sol and Claude Opus 5.

We report pass@$k$ for $k\in\{1,10\}$.
Pass@$k$ measures the probability that at least one of $k$ generated HDL implementations passes for a problem.
Following Chen et al.~\cite{chen2021codex} and VerilogEval~\cite{verilogeval}, we estimate pass@$k$ from the $n=10$ HDL implementations generated for each problem as
\[
  \mathrm{pass@}k=\frac{1}{|P|}\sum_{p\in P}
  \left(1-\frac{\binom{n-c_p}{k}}{\binom{n}{k}}\right),
\]
where $P$ is the problem set and $c_p$ is the number of generated HDL implementations that pass for problem $p$.

\subsection{Capability Differences Revealed by CPU-Eval}

As shown in Table~\ref{tab:results}, mean pass@1 across the seven LLMs is \SevenModelMeanPassOneGap{} percentage points lower on CPU-Eval than on VerilogEval v2 S2R.
Claude Opus 5 achieves \OpusVEOverallPassOne\% pass@1 on VerilogEval v2 S2R but only \OpusCPUOverallPassOne\% on CPU-Eval, showing that a high VerilogEval pass rate does not imply comparable ability to generate HDL for submodules in practical CPUs.
CPU-Eval reveals patterns hidden by VerilogEval's high pass rates, as described below.

\subsubsection{Sensitivity to Source-Code Length}

\begin{figure}[t]
  \centering
  \includegraphics[width=\columnwidth]{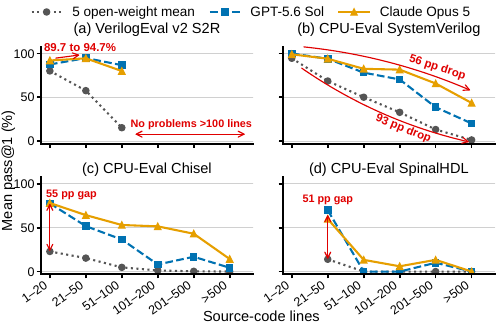}
  \caption{%
    Mean pass@1 by source-code length.
    (a) highlights the near-saturated mean across the two commercial models on VerilogEval v2 S2R.
    (b) illustrates how pass@1 decreases across models as SystemVerilog submodules in CPU-Eval become longer.
    (c) and (d) highlight differences in mean pass@1 between the five open-weight and two commercial models for Chisel and SpinalHDL submodules.}
  \label{fig:length-hdl-pass}
\end{figure}

Near-saturated scores on VerilogEval v2 S2R obscure how pass@1 changes with source-code length.
As shown in Fig.~\ref{fig:length-hdl-pass}(a), mean pass@1 across the two commercial models increases from 89.74\% at 1--20 lines to 94.66\% at 21--50 lines despite the increase in source-code length.
Both scores are near saturation, and most remaining failures in the 1--20-line range arise from defective prompts or tests of the kinds discussed in Section~\ref{sec:benchmark-quality}.
The apparent reversal therefore provides little information about the models' sensitivity to source-code length.

Figs.~\ref{fig:length-hdl-pass}(b) and (c) show an overall downward trend in mean pass@1 as source-code length increases for CPU-Eval's SystemVerilog and Chisel submodules.
The range above 500 lines in Fig.~\ref{fig:length-hdl-pass}(b) includes CPU-Eval's longest submodule at 3,176 lines, whereas the longest solution provided with a VerilogEval v2 S2R problem contains only 78 lines.
In both panels, each series is substantially lower in its longest populated range than in its shortest populated range.

Fig.~\ref{fig:length-hdl-pass}(b) reveals that pass@1 decreases differently across models as the number of source-code lines in a SystemVerilog submodule increases.
The mean for the five open-weight models nearly collapses from 94.55\% in the 1--20-line range to 1.38\% above 500 lines, a 93.17-percentage-point decrease.
GPT-5.6 Sol and Claude Opus 5 show smaller decreases of 79.79 and 55.55 percentage points, respectively, retaining pass@1 values of 20.21\% and 43.54\% above 500 lines.
CPU-Eval thus distinguishes models not only by their overall pass@1 but also by how well they generate submodules with more source-code lines.

ChipVerilog reports that LLMs generate functionally correct HDL less often as source-code length increases across modules from several hardware designs, including two small CPUs~\cite{chipverilog}.
CPU-Eval shows the same overall tendency across submodules drawn exclusively from six complete CPUs and further reveals that the decrease differs among LLMs and HDLs.

This sensitivity to source-code length has an implication for LLM-aided hardware design.
Modular design makes complex systems more manageable for human developers by decomposing them into smaller components that can be developed and tested independently~\cite{kravchuk2025modular}.
Our results suggest that the smaller submodules of fine-grained modular design can likewise make hardware-generation tasks more tractable for LLMs.

\subsubsection{Differences among HDLs}

Figs.~\ref{fig:length-hdl-pass}(b)--(d) reveal substantial differences among CPUs written in different HDLs even after grouping their submodules by source-code length.
In every source-code-length range shared by all three HDLs, the seven-model mean pass@1 is highest on the SystemVerilog CPUs and substantially lower on the Chisel and SpinalHDL CPUs.
For example, among submodules containing 51--100 lines, the two commercial models average 80.31\% on the SystemVerilog CPUs, 44.83\% on the Chisel CPUs, and 6.67\% on the SpinalHDL CPU.
Success on SystemVerilog submodules therefore does not characterize a model's ability to generate submodules for CPUs written in Chisel or SpinalHDL.

Figs.~\ref{fig:length-hdl-pass}(c) and (d) also show that this difficulty varies substantially across models.
For Chisel submodules containing 1--20 lines, the two commercial models outperform the five open-weight models by 55 percentage points.
For SpinalHDL submodules containing 21--50 lines, the difference is 51 percentage points.

One hypothesis is that models benefit from public information about an HDL.
As of September 19, 2026, Stack Overflow, a public question-and-answer website for programmers, contains 3,672 questions tagged \texttt{system-verilog}, 695 tagged \texttt{chisel}, and 13 tagged \texttt{spinalhdl}.
These question counts follow the same order as overall mean pass@1 across the three HDLs in Figs.~\ref{fig:length-hdl-pass}(b)--(d).
These results suggest that less widely used HDLs present an adoption barrier for LLM-aided hardware design.
LLMs may overcome this barrier: the commercial models substantially outperform the open-weight models on both Chisel and SpinalHDL.

\subsubsection{Differences Between CPUs Written in the Same HDL}

\begin{figure}[t]
  \centering
  \includegraphics[width=\columnwidth]{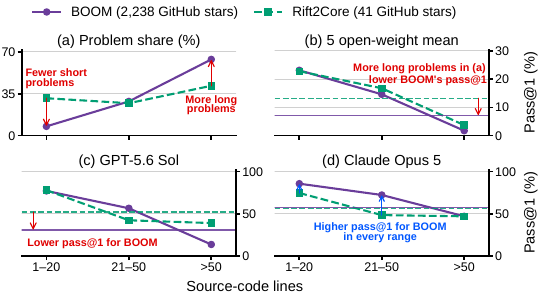}
  \caption{%
    BOOM and Rift2Core results by source-code length.
    (a) shows the share of problems in each range.
    (b)--(d) show mean pass@1 in each range, with horizontal lines marking pass@1 across all problems from each CPU.}
  \label{fig:boom-rift-length-comparison}
\end{figure}

Fig.~\ref{fig:boom-rift-length-comparison} compares BOOM and Rift2Core, two out-of-order CPUs written in Chisel.
Fig.~\ref{fig:boom-rift-length-comparison}(a) shows that BOOM has a larger share of problems above 50 lines and a smaller share of problems containing 1--20 lines than Rift2Core.
For the five open-weight models in Fig.~\ref{fig:boom-rift-length-comparison}(b), the two CPUs differ by at most 3 percentage points within each source-code-length range, but BOOM's overall pass@1 is 6.15 percentage points lower.
BOOM's larger share of long problems, where pass@1 is lower, therefore contributes to this overall difference.

Figs.~\ref{fig:boom-rift-length-comparison}(c) and (d) show different patterns for the two commercial models.
GPT-5.6 Sol, like the five open-weight models, has lower overall pass@1 on BOOM than on Rift2Core.
Claude Opus 5 differs: its pass@1 is higher on BOOM in every source-code-length range, and its overall pass@1 is nearly equal on the two CPUs.
As of September 19, 2026, BOOM has 2,238 GitHub stars and 529 forks, compared with 41 stars and 14 forks for Rift2Core.
Claude Opus 5 may therefore have encountered more BOOM-specific code, which could account for its higher pass@1 on BOOM within every range.

\FloatBarrier
\section{Conclusion}

This paper presented CPU-Eval, which evaluates 339 HDL-generation problems from six practical CPUs by integrating each generated submodule into its CPU and executing software.
Across seven LLMs, mean pass@1 is \SevenModelMeanPassOneGap{} percentage points lower on CPU-Eval than on VerilogEval v2 S2R, exposing capability differences hidden by VerilogEval's high pass rates.
Our evaluation revealed that pass@1 decreases as the number of source-code lines in a submodule increases, motivating the use of smaller submodules as LLM generation units.
The comparison among HDLs also revealed lower pass@1 for CPUs written in less widely used HDLs.
This result underscores the need for practical hardware-code-generation benchmarks such as CPU-Eval to assess an LLM's proficiency in generating code in the HDL designers use before deployment.

\section*{Acknowledgment}

OpenAI Codex and Anthropic Claude were used to assist with constructing the benchmark, conducting the experiments, and editing the manuscript.
The authors reviewed and verified all AI-assisted content.
This work is partly supported by JST Research and Development Program for Next-generation Edge AI Semiconductors, Grant Number JPMJES2511.

\IEEEtriggeratref{18}
\bibliographystyle{IEEEtran}
\bibliography{references}

\end{document}

%% file: problem-grid.tex
\newsavebox{\ProblemNameBox}
\newsavebox{\ProblemSingleLineBox}
\newsavebox{\ProblemLinesBox}
\newlength{\ProblemNameLimit}
\newlength{\ProblemValueShift}
\providecommand{\InstructionsColumnWidth}{8mm}
\providecommand{\ProblemLinesColumnWidth}{5.5mm}
\providecommand{\ProblemNameLeading}{7.2}
\providecommand{\ProblemRowGapExtra}{0pt}
\providecommand{\ProblemNameVAlign}{c}
\providecommand{\ProblemValueVAlign}{c}
\providecommand{\ProblemNameEndStrut}{}
\providecommand{\ProblemLineSkipLimit}{0pt}
\providecommand{\SetProblemValueShift}{\setlength{\ProblemValueShift}{0pt}}
\newcommand{\ProblemEntry}[3]{%
  \sbox{\ProblemLinesBox}{#2}%
  \setlength{\ProblemNameLimit}{%
    \dimexpr\linewidth-1.65mm-\InstructionsColumnWidth-\wd\ProblemLinesBox\relax}%
  \sbox{\ProblemNameBox}{%
    \parbox[\ProblemNameVAlign]{\ProblemNameLimit}{%
      \raggedright\fontsize{8}{\ProblemNameLeading}\selectfont
      \lineskiplimit=\ProblemLineSkipLimit
      \hangindent=0.3em\hangafter=1\strut\path{#1}\ProblemNameEndStrut}}%
  \sbox{\ProblemSingleLineBox}{%
    \fontsize{8}{\ProblemNameLeading}\selectfont\strut X}%
  \ifdim\ht\ProblemNameBox<8pt
    \ht\ProblemNameBox=\ht\ProblemSingleLineBox
  \else\ifdim\ht\ProblemNameBox<15pt
    \ht\ProblemNameBox=\dimexpr\ht\ProblemSingleLineBox+\ProblemNameLeading pt\relax
  \else\ifdim\ht\ProblemNameBox<22pt
    \ht\ProblemNameBox=\dimexpr\ht\ProblemSingleLineBox+\ProblemNameLeading pt+\ProblemNameLeading pt\relax
  \else
    \ht\ProblemNameBox=\dimexpr\ht\ProblemSingleLineBox+\ProblemNameLeading pt+\ProblemNameLeading pt+\ProblemNameLeading pt\relax
  \fi\fi\fi
  \SetProblemValueShift
  \noindent\hspace*{0.8mm}\hbox to \dimexpr\linewidth-1.5mm\relax{%
    \makebox[\dimexpr\linewidth-1.80mm-\ProblemLinesColumnWidth-\InstructionsColumnWidth\relax][l]{%
      \usebox{\ProblemNameBox}}%
    \hspace{0.25mm}%
    \raisebox{\dimexpr-0.5pt+\ProblemValueShift\relax}[0pt][0pt]{%
      \parbox[\ProblemValueVAlign]{\ProblemLinesColumnWidth}{\raggedleft\strut #2\par}}%
    \hspace{0.55mm}%
    \raisebox{\dimexpr-0.5pt+\ProblemValueShift\relax}[0pt][0pt]{%
      \parbox[\ProblemValueVAlign]{\InstructionsColumnWidth}{\raggedleft\strut #3\par}}%
    \hss}%
}
\newcommand{\ProblemColumn}[2][0pt]{%
  \begin{minipage}[t]{0.2\textwidth}\vspace{0pt}%
    \def\ProblemRowGap{\dimexpr#1+\ProblemRowGapExtra\relax}%
    \renewcommand{\\}[1][0pt]{\par\nointerlineskip\vspace{##1}}%
    #2%
  \end{minipage}%
}
\newcommand{\ProblemHeaders}{%
  \ProblemHeaderColumn\ProblemHeaderColumn\ProblemHeaderColumn%
  \ProblemHeaderColumn\ProblemHeaderColumn%
}
\providecommand{\ProblemLabel}{Problem}
\providecommand{\LinesLabel}{Lines}
\providecommand{\InstructionsLabel}{Insn.}
\providecommand{\ProblemHeaderStyle}{\footnotesize\bfseries}
\newcommand{\ProblemHeaderColumn}{%
  \begin{minipage}[t]{0.2\textwidth}\vspace{-1.6ex}%
    \hspace*{0.8mm}%
    \ProblemHeaderStyle
    \makebox[\dimexpr\linewidth-2.30mm-\ProblemLinesColumnWidth-\InstructionsColumnWidth\relax][l]{\ProblemLabel}%
    \hspace{0.25mm}\makebox[\ProblemLinesColumnWidth][r]{\LinesLabel}%
    \hspace{0.55mm}\makebox[\InstructionsColumnWidth][r]{\InstructionsLabel}%
    \par\vspace{-1.8ex}\smash{\hspace*{0.8mm}\rule{\dimexpr\linewidth-1.6mm\relax}{0.3pt}}%
  \end{minipage}%
}

\newcommand{\ProblemGrid}[1]{%
  \begin{minipage}[t]{\textwidth}\vspace{0pt}%
    \noindent\rule{\textwidth}{0.6pt}%
    \par\vspace{-0.65ex}%
    \renewcommand{\arraystretch}{0.78}%
    \ProblemHeaders%
    \par\vspace{-1.65ex}%
    #1%
    \par\vspace{0.4ex}\rule{\textwidth}{0.3pt}%
  \end{minipage}%
}
\newcommand{\BOOMRows}{%
  \ProblemColumn[0pt]{%
    \ProblemEntry{ALU}{50}{14.6k} \\[\ProblemRowGap]
    \ProblemEntry{ALUExeUnit}{225}{4.24k} \\[\ProblemRowGap]
    \ProblemEntry{ALUUnit}{186}{18.1k} \\[\ProblemRowGap]
    \ProblemEntry{AMOALU}{56}{3.62k} \\[\ProblemRowGap]
    \ProblemEntry{Arbiter}{17}{3.58k} \\[\ProblemRowGap]
    \ProblemEntry{BasicDispatcher}{19}{5.48k} \\[\ProblemRowGap]
    \ProblemEntry{BIMBranchPredictorBank}{106}{3.17k\smash{\textsuperscript{$\dagger$}}} \\[\ProblemRowGap]
    \ProblemEntry{BoomCore}{1430}{4.45k} \\[\ProblemRowGap]
    \ProblemEntry{BoomDuplicatedDataArray}{23}{1.21k} \\[\ProblemRowGap]
    \ProblemEntry{BoomFrontendModule}{670}{2.98k} \\[\ProblemRowGap]
    \ProblemEntry{BoomIOMSHR}{82}{1.53k} \\[\ProblemRowGap]
    \ProblemEntry{BoomMSHR}{354}{7.77k} \\[\ProblemRowGap]
    \ProblemEntry{BoomMSHRFile}{265}{7.77k} \\[\ProblemRowGap]
    \ProblemEntry{BoomNonBlockingDCacheModule}{502}{11.6k} \\[\ProblemRowGap]
    \ProblemEntry{BoomProbeUnit}{112}{$\dagger$} \\[\ProblemRowGap]
    \ProblemEntry{BoomRAS}{22}{$\dagger$} \\[\ProblemRowGap]
    \ProblemEntry{BoomTileModuleImp}{98}{2.42k} \\[\ProblemRowGap]
    \ProblemEntry{BoomWritebackUnit}{120}{$\dagger$} \\[\ProblemRowGap]
    \ProblemEntry{BranchDecode}{91}{3.17k\smash{\textsuperscript{$\dagger$}}} \\[\ProblemRowGap]
  }%
  \ProblemColumn[0pt]{%
    \ProblemEntry{BranchKillableQueue}{88}{2.26k} \\[\ProblemRowGap]
    \ProblemEntry{BranchMaskGenerationLogic}{76}{3.17k\smash{\textsuperscript{$\dagger$}}} \\[\ProblemRowGap]
    \ProblemEntry{BranchPredictor}{278}{3.17k\smash{\textsuperscript{$\dagger$}}} \\[\ProblemRowGap]
    \ProblemEntry{BreakpointUnit}{46}{$\dagger$} \\[\ProblemRowGap]
    \ProblemEntry{BTBBranchPredictorBank}{176}{2.13k\smash{\textsuperscript{$\dagger$}}} \\[\ProblemRowGap]
    \ProblemEntry{BundleBridgeNexus}{37}{$\dagger$} \\[\ProblemRowGap]
    \ProblemEntry{CompareRecFN}{41}{700} \\[\ProblemRowGap]
    \ProblemEntry{ComposedBranchPredictorBank}{35}{$\dagger$} \\[\ProblemRowGap]
    \ProblemEntry{CSRFile}{1212}{1.64k} \\[\ProblemRowGap]
    \ProblemEntry{DecodeUnit}{128}{8.45k} \\[\ProblemRowGap]
    \ProblemEntry{DivSqrtRecF64}{42}{274} \\[\ProblemRowGap]
    \ProblemEntry{DivSqrtRecF64_mulAddZ31}{66}{274} \\[\ProblemRowGap]
    \ProblemEntry{DivSqrtRecF64ToRaw_mulAddZ31}{664}{274} \\[\ProblemRowGap]
    \ProblemEntry{DivUnit}{25}{469} \\[\ProblemRowGap]
    \ProblemEntry{FAMicroBTBBranchPredictorBank}{131}{$\dagger$} \\[\ProblemRowGap]
    \ProblemEntry{FDivSqrtUnit}{145}{535} \\[\ProblemRowGap]
  }%
  \ProblemColumn[0pt]{%
    \ProblemEntry{FetchBuffer}{160}{595} \\[\ProblemRowGap]
    \ProblemEntry{FetchTargetQueue}{268}{1.27k\smash{\textsuperscript{$\dagger$}}} \\[\ProblemRowGap]
    \ProblemEntry{FMADecoder}{31}{1.27k} \\[\ProblemRowGap]
    \ProblemEntry{ForwardingAgeLogic}{44}{243} \\[\ProblemRowGap]
    \ProblemEntry{FpPipeline}{249}{1.36k} \\[\ProblemRowGap]
    \ProblemEntry{FPToFP}{59}{1.34k} \\[\ProblemRowGap]
    \ProblemEntry{FPToInt}{73}{4.03k} \\[\ProblemRowGap]
    \ProblemEntry{FPU}{81}{2.31k} \\[\ProblemRowGap]
    \ProblemEntry{FPUExeUnit}{136}{1.75k} \\[\ProblemRowGap]
    \ProblemEntry{FPUFMAPipe}{36}{1.27k} \\[\ProblemRowGap]
    \ProblemEntry{FPUUnit}{21}{641} \\[\ProblemRowGap]
    \ProblemEntry{HellaCacheArbiter}{79}{$\dagger$} \\[\ProblemRowGap]
    \ProblemEntry{ICacheModule}{250}{$\dagger$} \\[\ProblemRowGap]
    \ProblemEntry{IntToFP}{44}{1.88k} \\[\ProblemRowGap]
    \ProblemEntry{IntToFPUnit}{45}{536} \\[\ProblemRowGap]
    \ProblemEntry{IntXbar}{18}{$\dagger$} \\[\ProblemRowGap]
    \ProblemEntry{IssueSlot}{227}{1.39k} \\[\ProblemRowGap]
    \ProblemEntry{IssueUnitCollapsing}{100}{$\dagger$} \\[\ProblemRowGap]
    \ProblemEntry{L1MetadataArray}{26}{$\dagger$} \\[\ProblemRowGap]
    \ProblemEntry{LoopBranchPredictorBank}{195}{$\dagger$} \\[\ProblemRowGap]
  }%
  \ProblemColumn[0pt]{%
    \ProblemEntry{LoopBranchPredictorColumn}{141}{$\dagger$} \\[\ProblemRowGap]
    \ProblemEntry{LSU}{1451}{4.45k} \\[\ProblemRowGap]
    \ProblemEntry{MaxPeriodFibonacciLFSR}{48}{$\dagger$} \\[\ProblemRowGap]
    \ProblemEntry{MemAddrCalcUnit}{73}{6.2k\smash{\textsuperscript{$\dagger$}}} \\[\ProblemRowGap]
    \ProblemEntry{Mul54}{43}{274} \\[\ProblemRowGap]
    \ProblemEntry{MulAddRecFNPipe}{62}{1.27k} \\[\ProblemRowGap]
    \ProblemEntry{MulAddRecFNToRaw_postMul}{126}{652} \\[\ProblemRowGap]
    \ProblemEntry{MulAddRecFNToRaw_preMul}{96}{652} \\[\ProblemRowGap]
    \ProblemEntry{MulDiv}{144}{4.78k} \\[\ProblemRowGap]
    \ProblemEntry{NBDTLB}{341}{$\dagger$} \\[\ProblemRowGap]
    \ProblemEntry{NullLocalBranchPredictorBank}{5}{$\dagger$} \\[\ProblemRowGap]
    \ProblemEntry{NullPrefetcher}{5}{$\dagger$} \\[\ProblemRowGap]
    \ProblemEntry{PipelinedMultiplier}{27}{2.94k} \\[\ProblemRowGap]
    \ProblemEntry{PipelinedMulUnit}{18}{2.94k} \\[\ProblemRowGap]
    \ProblemEntry{PMPChecker}{48}{$\dagger$} \\[\ProblemRowGap]
    \ProblemEntry{PredRenameStage}{49}{290\smash{\textsuperscript{$\dagger$}}} \\[\ProblemRowGap]
    \ProblemEntry{PTW}{476}{$\dagger$} \\[\ProblemRowGap]
  }%
  \ProblemColumn[0pt]{%
    \ProblemEntry{Queue}{89}{2.03k\smash{\textsuperscript{$\dagger$}}} \\[\ProblemRowGap]
    \ProblemEntry{RecFNToRecFN}{46}{536} \\[\ProblemRowGap]
    \ProblemEntry{RegisterFileSynthesizable}{74}{6.29k} \\[\ProblemRowGap]
    \ProblemEntry{RegisterRead}{186}{6.29k} \\[\ProblemRowGap]
    \ProblemEntry{RegisterReadDecode}{51}{7.77k} \\[\ProblemRowGap]
    \ProblemEntry{RenameBusyTable}{48}{6.29k} \\[\ProblemRowGap]
    \ProblemEntry{RenameFreeList}{78}{1.57k\smash{\textsuperscript{$\dagger$}}} \\[\ProblemRowGap]
    \ProblemEntry{RenameMapTable}{87}{6.66k} \\[\ProblemRowGap]
    \ProblemEntry{RenameStage}{195}{1.57k\smash{\textsuperscript{$\dagger$}}} \\[\ProblemRowGap]
    \ProblemEntry{Rob}{667}{2.22k\smash{\textsuperscript{$\dagger$}}} \\[\ProblemRowGap]
    \ProblemEntry{RoundAnyRawFNToRecFN}{242}{300} \\[\ProblemRowGap]
    \ProblemEntry{RoundRawFNToRecFN}{26}{300} \\[\ProblemRowGap]
    \ProblemEntry{RVCExpander}{15}{391} \\[\ProblemRowGap]
    \ProblemEntry{TageBranchPredictorBank}{163}{1.11k\smash{\textsuperscript{$\dagger$}}} \\[\ProblemRowGap]
    \ProblemEntry{TageTable}{160}{1.11k\smash{\textsuperscript{$\dagger$}}} \\[\ProblemRowGap]
    \ProblemEntry{TLB}{341}{$\dagger$} \\[\ProblemRowGap]
    \ProblemEntry{TLXbar}{47}{$\dagger$} \\[\ProblemRowGap]
    \ProblemEntry{UOPCodeFDivDecoder}{38}{535} \\[\ProblemRowGap]
    \ProblemEntry{UOPCodeFPUDecoder}{91}{6.64k} \\[\ProblemRowGap]
  }%
}%